\documentclass{article}

\RequirePackage[T1]{fontenc}
\RequirePackage[utf8]{inputenc}

\usepackage{amsmath}
\usepackage{amssymb}
\usepackage{graphicx}
\usepackage{hyphenat}
\usepackage{microtype}
\usepackage{multicol}
\usepackage{musicography}
\usepackage{siunitx}
\usepackage{spconf}
\usepackage{stackengine}
\usepackage{stmaryrd}
\usepackage[caption=false,font=footnotesize]{subfig}
\usepackage{todonotes}
\usepackage{url}
\usepackage{xspace}
\newcommand*{\eg}{e.g.,\@\xspace}
\newcommand*{\etc}{etc.\@\xspace}
\newcommand*{\ie}{i.e.,\@\xspace}

\DeclareMathOperator{\median}{median}

\title{Learning the helix topology of musical pitch}

\name{
Vincent Lostanlen$^{\star \dagger}$
\quad
Sripathi Sridhar$^{\dagger}$
\quad
Brian McFee$^{\dagger}$
\quad
Andrew Farnsworth$^{\star}$
\quad
Juan Pablo Bello$^{\dagger}$
\thanks{This work is partially supported by NSF award 1633259 (BirdVox).}}

\address{
$^{\star}$ Cornell Lab of Ornithology, Cornell University, Ithaca, NY, USA \\
$^{\dagger}$ Music and Audio Research Laboratory, New York University, New York, NY, USA}

\begin{document}
\ninept
\maketitle

%
\begin{abstract}
To explain the consonance of octaves, music psychologists represent pitch as a helix where azimuth and axial coordinate correspond to pitch class and pitch height respectively.
This article addresses the problem of discovering this helical structure from unlabeled audio data.
We measure Pearson correlations in the constant-$Q$ transform (CQT) domain to build a $K$-nearest neighbor graph between frequency subbands.
Then, we run the Isomap manifold learning algorithm to represent this graph in a three-dimensional space in which straight lines approximate graph geodesics.
Experiments on isolated musical notes demonstrate that the resulting manifold resembles a helix which makes a full turn at every octave.
A circular shape is also found in English speech, but not in urban noise.
We discuss the impact of various design choices on the visualization: instrumentarium, loudness mapping function, and number of neighbors $K$.
\end{abstract}

%

\section{Introduction}
\label{sec:intro}
\begin{figure}
\centering
\includegraphics[height=3cm]{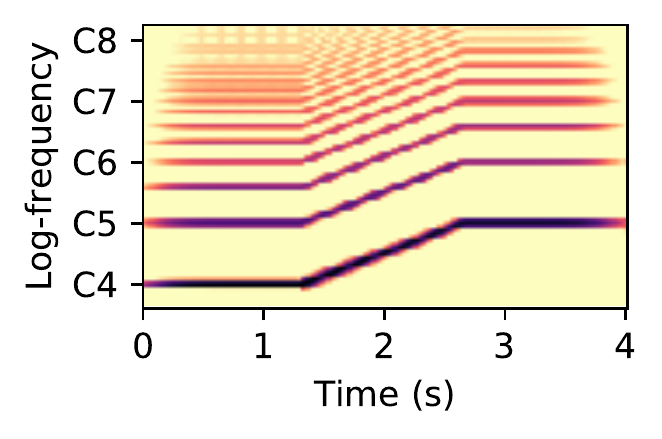}
\includegraphics[height=3cm]{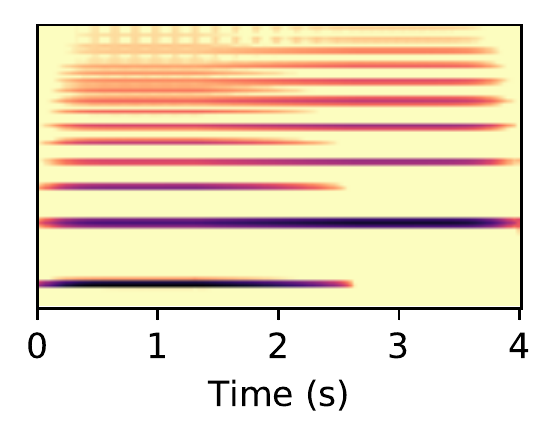}
\vspace*{-2mm}
\caption{Two continuous trajectories in pitch space, either by octave glissando (left) or by attenuation of odd-numbered partials (right). Darker shades indicate larger magnitudes of the constant-$Q$ transform. Axes $t$ and $\gamma$ respectively correspond to time and log-frequency. Vertical ticks denote octaves. See Section \ref{sec:intro} for details.}
\label{fig:odd-numbered-partials}
\end{figure}

%
Listening to a sequence of two pure tones elicits a sensation of pitch going ``up'' or ``down'', correlating with changes in fundamental frequency ($f_0$).
However, contrary to pure tones, natural pitched sounds contain a rich spectrum of components in addition to $f_0$.
Neglecting inharmonicity, these components are tuned to an ideal Fourier series whose modes resonate at integer multiples of the fundamental: $2f_0$, $3f_0$, and so forth.
By the change of variable $f_0^\prime = 2f_0$, it appears that all even-numbered partials $2f_0=f_0^{\prime} $, $4f_0=2 f_0^{\prime}$, $6f_0=3 f_0^{\prime}$, and so forth make up the whole Fourier series of a periodic signal whose fundamental frequency is $f_0^\prime$.
Figure \ref{fig:odd-numbered-partials} illustrates, in the case of a synthetic signal with perfect harmonicity, that filtering out all odd-numbered partials $(2p+1)f_0$ for integer $p\geq0$ results in a perceived pitch that morphs from $f_0$ to $2f_0$, \ie{} up one octave \cite{deutsch2008jasa}.

Such ambivalence brings about well-known auditory paradoxes: tones that have a definite pitch class but lack a pitch register \cite{deutsch2010acoustics}; glissandi that seem to ascend or descend endlessly \cite{risset1969jasa}; and tritone intervals whose pitch directionality depends on prior context \cite{pelofi2017philtrans}.
To explain them, one may roll up the frequency axis onto a spiral or helix which makes a full turn at each octave, thereby aligning power-of-two harmonics onto the same radii.

%
The consonance of octaves has implications in several disciplines.
In music theory, it allows for pitches to be grouped into pitch classes: \eg{} in European solf\`ege, \emph{do re mi} \etc{} eventually ``circle back'' to \emph{do}.
In ethnomusicology, it transcends the boundaries between cultures, to the point of being held as a nearly universal attribute of music \cite{burns1999chapter}.
In neurophysiology, it explains the functional organization of the central auditory cortex \cite{warren2003pnas}.
In music information research, it motivates the design of chroma features, \ie{} a representation of harmonic content that is meant to be equivariant to parallel chord progressions but invariant to chord inversions \cite[chapter 5]{muller2015fundamentals}.

%
Despite the wealth of evidence for the crucial role of octaves in music, there is, to this day, no data-driven criterion for assessing whether a given audio corpus exhibits a property of octave equivalence.
Rather, the disentanglement of pitch chroma and pitch height in time--frequency representations relies on domain-specific knowledge about music \cite{tymoczko2006science}.
More generally, although the induction of priors on the topology of pitch is widespread in symbolic music analysis \cite{bigo2016book}, few of them directly apply to audio signal processing \cite{chuan2005icme}.

Yet, in recent years, the systematic use of machine learning methods has progressively reduced the need for domain-specific knowledge in several other aspects of auditory perception, including mel-frequency spectrum \cite{zeghidour2018icassp} and adaptive gain control \cite{wang2017trainable}.
It remains to be known whether octave equivalence can, in turn, be discovered by a machine learning algorithm, instead of being engineered ad hoc.
Furthermore, it is unclear whether octave equivalence is an exclusive characteristic of music or whether it may extend to other kinds of sounds, such as speech or environmental soundscapes.

%
In this article, we conduct an unsupervised manifold learning experiment, inspired by the protocol of ``learning the 2-D topology of images'' \cite{leroux2008neurips}, to visualize the helix topology of constant-$Q$ spectra.
Starting from a dataset of isolated notes from various musical instruments, we run the Isomap algorithm to represent each of these frequency subbands as a dot in a 3-D space, wherein spatial neighborhoods denote high correlation in loudness.
Contrary to natural images, we find a mismatch between the physical dimensionality of natural acoustic spectra (\ie{} 1-D) and their statistical dimensionality (\ie{} 3-D or greater).

The companion website of this paper\footnote{Companion website: \url{https://github.com/BirdVox/lostanlen2020icassp}} contains a Python package to visualize octave equivalence in audio data.

\begin{figure*}
  \includegraphics[width=\textwidth]{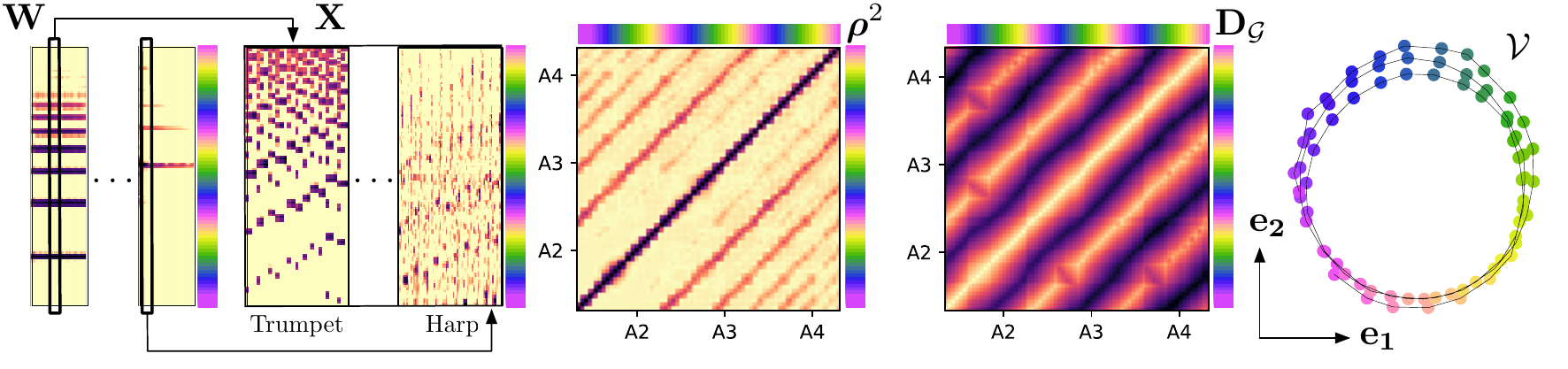}
  \caption{Functional diagram of the proposed method, comprising: constant-$Q$ transform $\mathbf{W}$, data matrix $\mathbf{X}$, extraction of Pearson correlations $\boldsymbol{\rho}^2$, shortest path distance matrix $\mathbf{D}_{\mathcal{G}}$, and Isomap eigenbasis $\mathbf{e}$ for the set of vertices $\mathcal{V}$.
  Darker shades in $\boldsymbol{\rho}^{2}$ and $\mathbf{D}_{\mathcal{G}}$ indicate larger absolute values of the Pearson correlation and distance respectively.
  The hue of colored dots and the solid grey line respectively denote pitch chroma and pitch height.
  The first three dimensions in the Isomap embedding explain 36\%, 35\%, and 9\% of the total variance in $\mathbf{D}_{\mathcal{G}}$ respectively.
  Note that our method is unsupervised: neither pitch chroma nor pitch height are directly supplied to the Isomap manifold learning algorithm.
  See Section \ref{sec:isomap} for details.}
  \label{fig:functional-diagram}
\end{figure*}

\section{Isomap embedding of subband correlations}
\label{sec:isomap}

\subsection{Constant-$Q$ transform and loudness mapping}
Given a corpus of audio signals $\boldsymbol{x}_1 \ldots \boldsymbol{x}_N \in\mathbf{L}^2(\mathbb{R})$ in arbitrary order, we define their constant-$Q$ transforms (CQT) as the convolutions with wavelets $t\mapsto 2^\gamma \boldsymbol{\psi}(2^\gamma t)$, where $2^{-\gamma}$ is a scale parameter:
\begin{equation}
    \mathbf{W}\boldsymbol{x}_n:
    (t, \gamma) \mapsto 2^\gamma
    \int_{-\infty}^{+\infty}
    \boldsymbol{x}_n(t^\prime )
    \boldsymbol{\psi}(2^{\gamma} (t-t^\prime))
    \,\mathrm{d}t.
\end{equation}
The wavelet filterbank in the operator $\mathbf{W}$ covers a range of $J$ octaves.
In the case of musical notes, we define a region of interest $t_n$ in each $\boldsymbol{x}_n$ as the frame of highest short-term energy at a scale of \SI{93}{\milli\second}.
We compute their scalogram representation as the vector of CQT modulus responses at some $t=t_n$, for discretized values $u = 1 + Q \gamma$:
\begin{equation}
    \mathbf{X}[n,u] =
    \left\vert \mathbf{W}\boldsymbol{x}_n\left(t_n, \frac{u-1}{Q}\right) \right\vert,
\end{equation}
where the integer $u$ ranges from $1$ to $QJ$.
Then, we apply a pointwise logarithmic compression $\Lambda$ to map each magnitude coefficient $\mathbf{X}[n,u]$ onto a decibel-like scale, and clip it to \SI{-100}{\deci\bel}:
\begin{equation}
    \Lambda(\mathbf{X})[n,u] =
    \max(-100, 10 \log_{10} \mathbf{\mathbf{X}}[n,u]).
    \label{eq:Lambda}
\end{equation}
Unless stated otherwise, we set $\vert \boldsymbol{\psi} \vert$ to a Hann window, the quality factor $Q$ to $24$, and the number of octaves $J$ to 3 in the following.
We compute constant-$Q$ transforms with librosa v0.6.1 \cite{mcfee2018librosa}.
Section \ref{sub:varying-Lambda} will discuss the effect of alternative choices for $\Lambda$.

\subsection{Pearson autocorrelation of log-magnitude spectra}

In accordance with \cite{leroux2008neurips}, we express the similarity between two features $u$ and $v$ in terms of their squared Pearson correlation $\boldsymbol{\rho}^2 [u,v]$.
We begin by recentering each feature to null mean, yielding the matrix
\begin{equation}
    \mathbf{Y}[n,u] =
    \Lambda(\mathbf{X})[n,u] -
    \dfrac{1}{N}
    \sum_{n=1}^{N} \Lambda(\mathbf{X})[n,u],
\end{equation}
and then compute squared cosine similarities on all pairs $(u,v)$:
\begin{equation}
    \boldsymbol{\rho}^2[u,v] =
    \dfrac{(\sum_{n=1}^{N} \mathbf{Y}[n,u] \mathbf{Y}[n,v])^2}{
    (\sum_{n=1}^{N} \mathbf{Y}^2[n,u]) \times
    (\sum_{n=1}^{N} \mathbf{Y}^2[n,v])}.
\end{equation}
Let $\mathcal{V}$ be the set of all $QJ$ features $u$ in $\mathbf{Y}$.
Adopting a manifold learning perspective, we may regard $\boldsymbol{\rho}^2$ as the values of a standard Gaussian kernel $\kappa:\mathcal{V}\times\mathcal{V}\longrightarrow\mathbb{R}$.
Inverting the identity $\boldsymbol{\rho}^2[u,v] = \kappa(u, v) = \exp(-2 \mathbf{D}_{\boldsymbol{\rho^2}}^2[u,v])$ leads to a pseudo-Euclidean distance
\begin{equation}
    \mathbf{D}_{\boldsymbol{\rho}^2}[u,v] =
    \sqrt{- \frac{1}{2} \log \boldsymbol{\rho}^2 [u,v]}.
\end{equation}

\subsection{Shortest path distance on the $K$-nearest neighbor graph}

Following the Isomap manifold learning algorithm \cite{tenenbaum2000science}, we compute the $K$ nearest neighbors of each feature $u\in\mathcal{V}$ as the set $\mathcal{N}_K (u)$ of $K$ features $v\neq u$ that minimize the distance $\mathbf{D}_{\boldsymbol{\rho^2}}[u,v]$, \ie{} maximize the squared Pearson correlation $\boldsymbol{\rho}^2[u,v]$.
We construct a $K$-nearest graph $\mathcal{G}$ whose vertices are $\mathcal{V}$ and whose adjacency matrix $\mathbf{A}[u,v]$ is equal to $\mathbf{D}_{\boldsymbol{\rho}}[u,v]$ if $v\in\mathcal{N}_K (u)$ or if $u\in\mathcal{N}_K  (v)$, and infinity otherwise.
We run Dijkstra's algorithm on $\mathcal{G}$ to measure geodesics on the manifold induced by $\mathbf{A}$. These geodesics yield a shortest path distance function over $\mathcal{V}^2$:
\begin{equation}
\mathbf{D}_{\mathcal{G}}[u,v] =
\left\{
\begin{array}{c}
\mathbf{D}_{\boldsymbol{\rho}^{2}}[u,v]
\text{ if }v\in\mathcal{N}_{K}(u)\text{ or }u\in\mathcal{N}_{K}(v)
\\
\displaystyle{\min_{z}}\;
(\mathbf{D}_{\mathcal{G}}[u,z]+\mathbf{D}_{\mathcal{G}}[z,v])
\text{ otherwise.}
\end{array}
\right.
\end{equation}
If $\mathcal{G}$ has more than one connected component, then $\mathbf{D}_\mathcal{G}$ is infinite over pairs of mutually unreachable vertices.
On some datasets, $\boldsymbol{\rho}$ may lead to a disconnected $K$-nearest neighbor graph $\mathcal{G}$, especially for small $K$, thereby causing numerical aberrations in Isomap.
However, we make sure that the effective bandwidth of the wavelet $\widehat{\boldsymbol{\psi}}$ is large enough in comparison with $Q$ to yield strong correlations $\boldsymbol{\rho}[u, u+1]$ for all $u$, and thus $u\in\mathcal{N}_K (u+1)$.
With this caveat in mind, we postulate that $\mathbf{D}_{\mathcal{G}}$ is finite in all of the following.
We set $K=3$ in the following unless stated otherwise.

\subsection{Classical multidimensional scaling and 3-D embedding}
Let $\mathbf{S}_{\mathcal{G}}[u,v] = \mathbf{D}_{\mathcal{G}}[u,v]^2$.
Classical multidimensional scaling (MDS) diagonalizes $\boldsymbol{\tau}_\mathcal{G} =-\frac{1}{2} (\mathbf{H}\mathbf{S}_\mathcal{G}\mathbf{H})$ where $\mathbf{H}[u,v] = \boldsymbol{\delta}(u-v) - \frac{1}{QJ}$ \cite{torgerson1952psychometrika}.
We denote by $\mathbf{e}_m$ and $\lambda_m$ the respective eigenvectors and eigenvalues of $\boldsymbol{\tau}_{\mathcal{G}}$, satisfying $\boldsymbol{\tau}_{\mathcal{G}}\mathbf{e}_m = \lambda_m \mathbf{e}_m$.
We rank eigenvalues in decreasing order, without loss of generality.
Lastly, we display the Isomap embedding as a 3-D scatter plot with Cartesian coordinates $\mathbf{e}[u] = (\mathbf{e_1}[u], \mathbf{e_2}[u], \mathbf{e_3}[u])$ for every vertex $u \in \mathcal{V}$.
We compute Isomap using scikit-learn v0.21.3 \cite{pedregosa2011scikitlearn}.

\begin{figure}
\centering
\subfloat[Trumpet in C only.]{
  \includegraphics[width=0.45\linewidth]{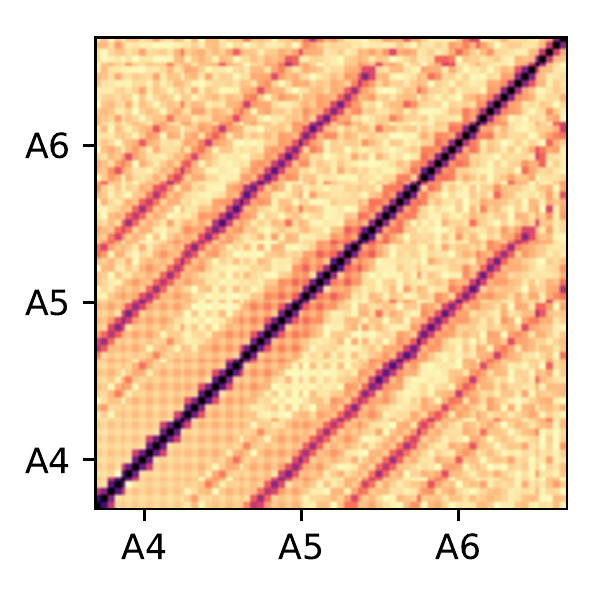}
  \includegraphics[width=0.45\linewidth]{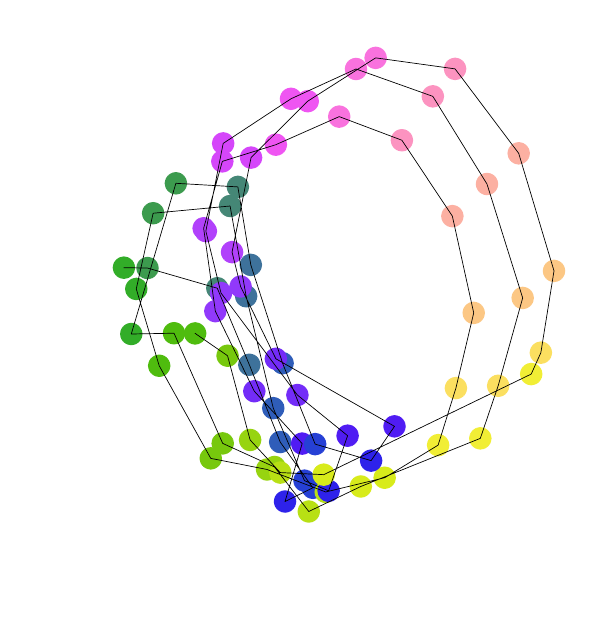}}
\\
\subfloat[Harp only.]{
  \includegraphics[width=0.45\linewidth]{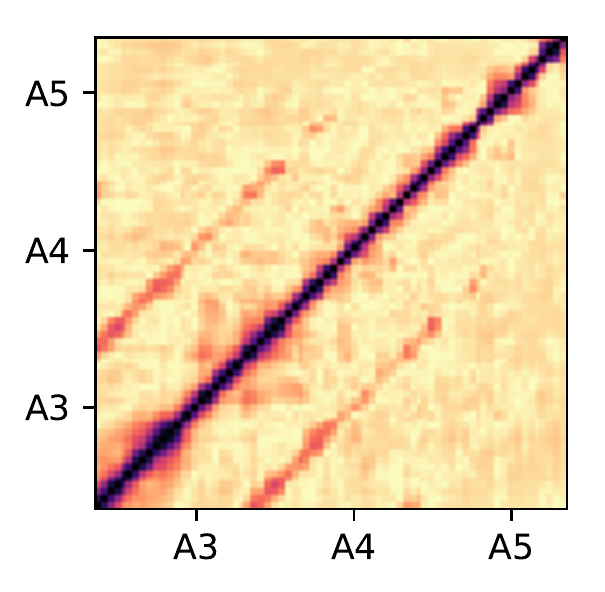}
  \includegraphics[width=0.45\linewidth]{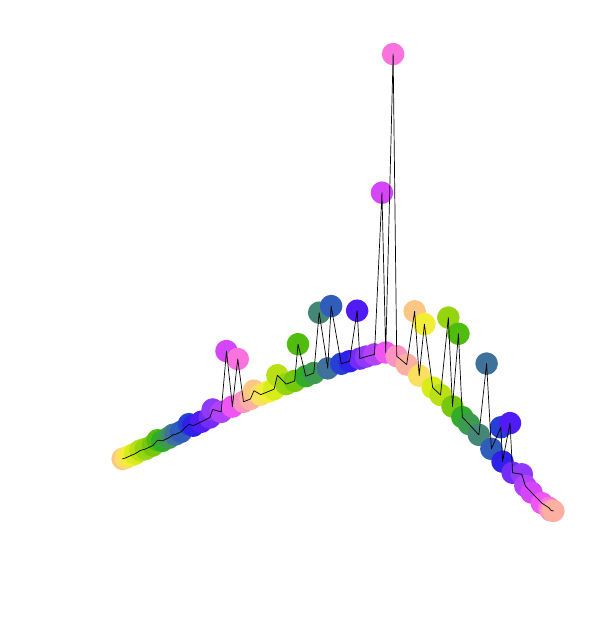}}
\caption{
Pearson correlation matrices $\boldsymbol{\rho}^2$ (left) and Isomap embeddings (right) for two instruments in the SOL dataset: Trumpet in C (a) and Harp (b).
In both cases, the number of neighbors is set to $K=3$ and the loudness mapping $\Lambda$ is logarithmic.
See Section \ref{sub:varying-instruments} for details.}
\label{fig:varying-instruments}
\end{figure}

\section{Experiments with musical sounds}
\label{sec:experiments-music}

\subsection{Main protocol}

We extract $\mathbf{X}$ from the isolated musical notes played by eight instruments in the SOL dataset \cite{ballet1999jim}: accordion, alto saxophone, bassoon, flute, harp, trumpet in C, and cello.
For each of these instruments, we include three levels of intensity dynamics: \emph{pp}, \emph{mf}, and \emph{ff}.
We include all available pitches in the tessitura ($\min=\mathrm{B_0}$, $\median=\textrm{E}_4$, $\max=\textrm{C}\musSharp_8$).
We exclude extended playing techniques, because some of them may lack a discernible pitch class \cite{lostanlen2018dlfm}; thus resulting in a total of $N=1212$ audio recordings for the \emph{ordinario} technique.
We hypothesize that cross-correlations across octaves are weaker than cross-correlations along the log-frequency axis.
Because the helix topology of musical pitch relies on both kinds of correlation, we set the number of neighbors to $K=3$.

Figure \ref{fig:functional-diagram} illustrates our protocol and main finding.
Isomap produces a quasi-perfect cylindrical manifold in dimension three.
We color each dot $\mathbf{e}[u]$ according to a hue of $\boldsymbol{\theta}(u) = \frac{2\pi}{Q} (u\mod Q)$ radians.
Furthermore, we draw a segment between each $\mathbf{e}[u]$ and its upper adjacent subband $\mathbf{e}[u+1]$.
Once these visual elements are included in the display of the scatter plot $(\mathbf{e}[u])_u$, the cylindrical manifold appears to coincide with the Drobisch-Shepard helix in music psychology \cite{shepard1964jasa,lerdahl2004book}.
Indeed, hues $\boldsymbol{\theta}[u]$ appear to align on the same radii, whereas the grey line $\gamma$ grows monotonically with $\mathbf{e_3}$.
This result demonstrates that, even without prior knowledge about the perception of musical pitch, it is possible to discover octave equivalence in a purely data-driven fashion, by computing the graph of greatest cross-correlations between CQT magnitudes in a corpus of isolated musical notes.

\subsection{Varying the instrumentarium \label{sub:varying-instruments}}
We reproduce the main protocol on subsets of the SOL dataset, involving a single instrument at once.
The bright timbre of brass instruments correlates with relatively loud high-order partials \cite{poirson2005jasa}, resulting in large octave equivalence and a helical topology (see Figure \ref{fig:varying-instruments} (a)).
In contrast, harp tones carry little energy at twice the fundamental; yet, they induce sympathetic resonance along the soundboard, which affects nearby strings predominantly \cite{lecarrou009actaacustica}.
These two phenomena in combination favor semitone correlations over octave correlations, resulting in a rectilinear topology (see Figure \ref{fig:varying-instruments} (a)).

\subsection{Varying the number of neighbors $K$ \label{sub:varying-neighbors}}
\begin{figure}
\centering
\subfloat[$K=2$ neighbors.]{
\includegraphics[width=0.45\linewidth]{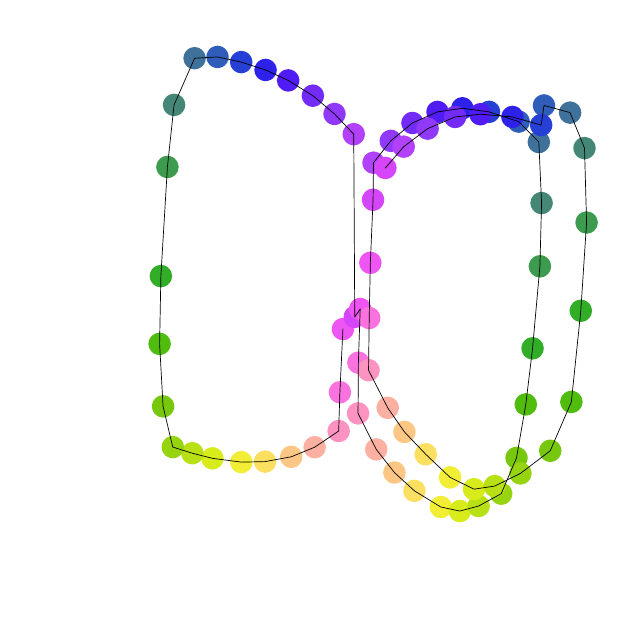}}
\subfloat[$K=5$ neighbors.]{
\includegraphics[width=0.45\linewidth]{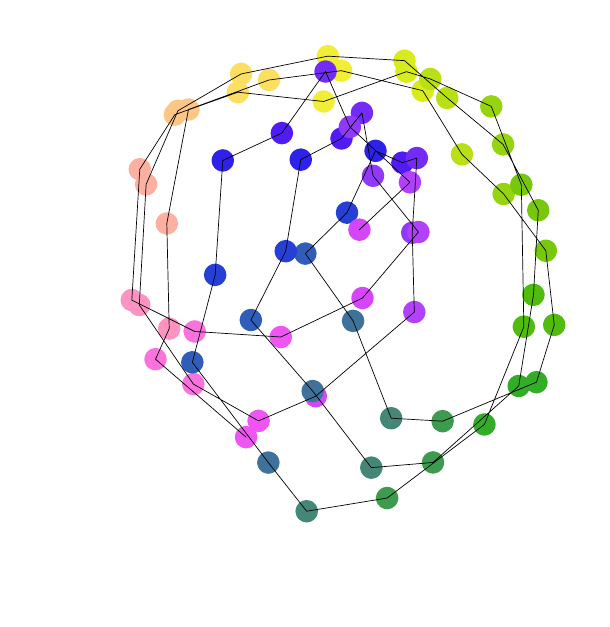}}
\caption{
Isomap embeddings for varying $K$-nearest neighbor graphs: $K=2$ (left) and $K=5$ (right).
In both cases, the Pearson correlation matrix $\boldsymbol{\rho}^2$ results from all instruments in the SOL dataset and the loudness mapping $\Lambda$ is logarithmic.
See Section \ref{sub:varying-neighbors} for details.}
\label{fig:varying-neighbors}
\end{figure}

We reproduce the main protocol with varying $K$-nearest neighbor graphs.
Setting $K=2$ results in multiple lobes, each corresponding to an octave, and connected at a single pitch class (see Figure \ref{fig:varying-neighbors}).
This confirms our hypothesis that large semitone correlations outnumber large octave correlations.
Conversely, setting $K=5$ results in a topology that is more intricate than the Drobisch-Shepard helix, involving correlations across perfect fourths and fifths..

\begin{figure}
\centering
\subfloat[Linear loudness mapping function.]{
\includegraphics[width=0.45\linewidth]{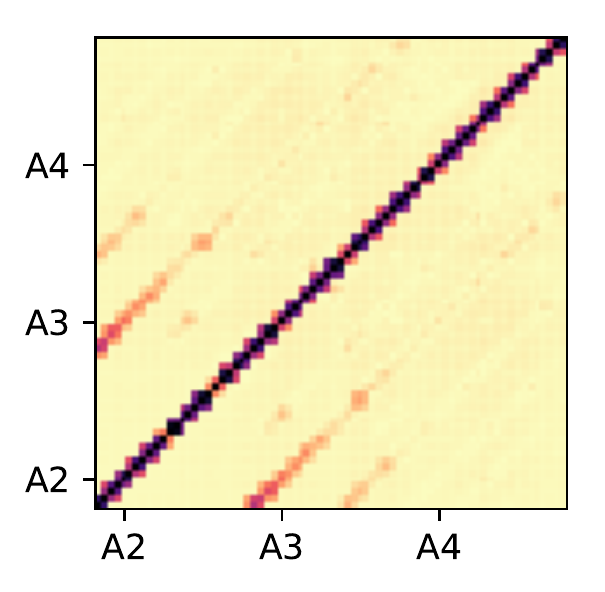}
\includegraphics[width=0.45\linewidth]{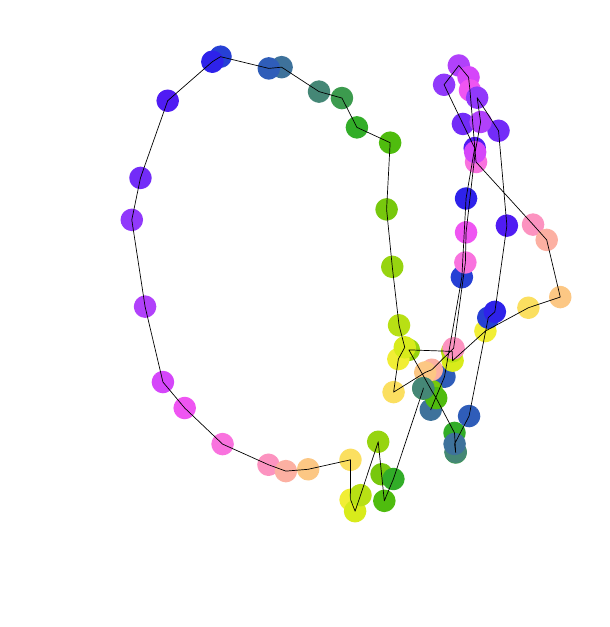}}
\\
\subfloat[Cubic root loudness mapping function.]{
\includegraphics[width=0.45\linewidth]{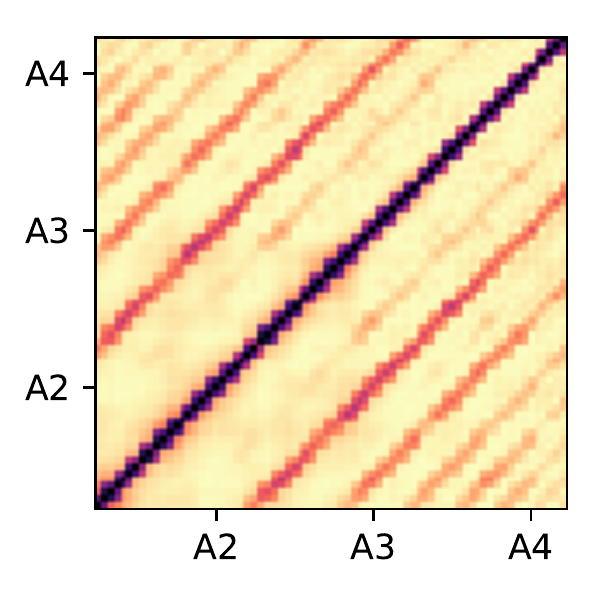}
\includegraphics[width=0.45\linewidth]{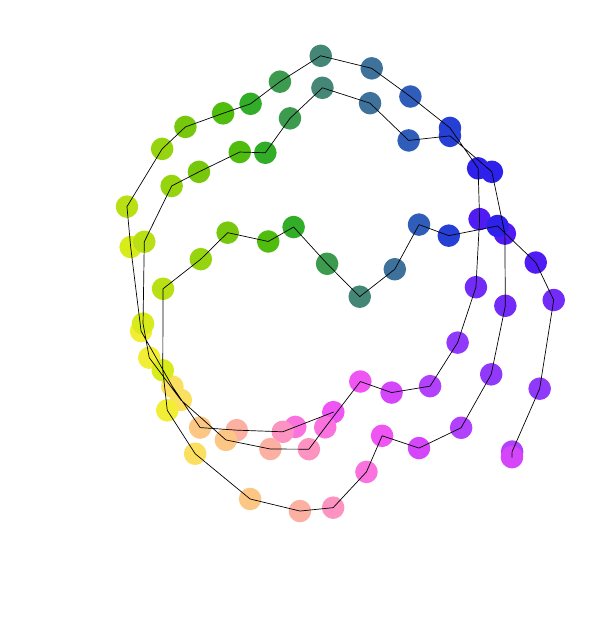}}
\caption{
Pearson correlation matrices $\boldsymbol{\rho}^2$ (left) and Isomap embeddings (right) for cubic root (a) and linear (b) loudness mappings.
In both cases, $\boldsymbol{\rho}^2$ results from all instruments in the SOL dataset and the number of neighbors is set to $K=3$.
See Section \ref{sub:varying-Lambda} for details.}
\label{fig:varying-loudness}
\end{figure}


\subsection{Varying the loudness mapping function $\Lambda$\label{sub:varying-Lambda}}
We reproduce the main protocol with varying loudness mappings $\Lambda$ (see Figure \ref{fig:varying-loudness}).
On one hand, setting $\Lambda$ to the identity function yields a $K$-nearest neighbor graph in which octave correlations are numerically negligible, except in the lower register.
This results in an Isomap embedding which is circular in the bottommost octave and irregular in the topmost octaves.
On the other hand, setting $\Lambda$ to the cubic root function yields a helical topology.
This experiment demonstrates the need for nonlinear loudness compression in the protocol; in contrast with \cite{leroux2008neurips}, which relied on raw grayscale intensities in handwritten digits.

\section{Extension to speech and urban sounds}
\label{sec:experiments-nonmusic}

\subsection{Experiment with speech data}
We analyze the North Texas vowel database (NTVOW), which contains utterances of 12 English vowels from 50 American speakers, including children aged three to seven as well as male and female adults \cite{assmann2000jasa}; resulting in a total of $N=9570$ audio recordings. As seen in Figure \ref{fig:non-music-data} (a), which includes the data from all age groups, there is some notion of pitch circularity seen in the mid-frequency range, but not so in the low and high-frequency ranges. This is because the distribution of $f_0$ in human speech is polarized around certain pitch classes, rather than uniform over the chromatic scale. 

\subsection{Experiment with environmental audio data}
\begin{figure}
\centering
\subfloat[Spoken English vowels (NTVOW dataset).]{
  \includegraphics[width=0.45\linewidth]{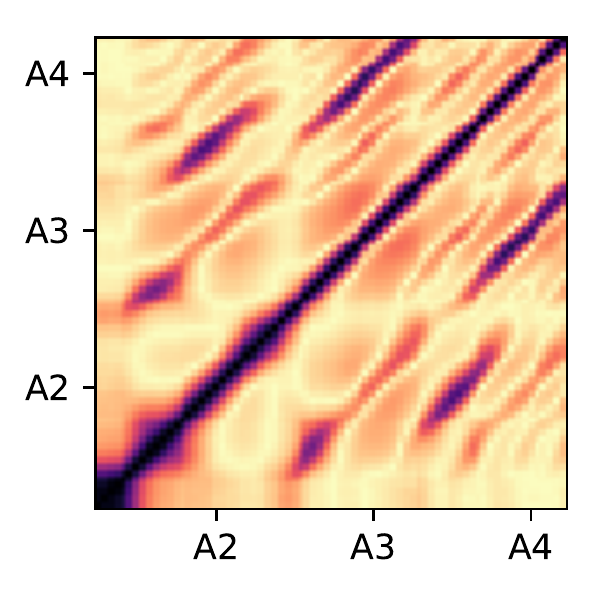}
  \includegraphics[width=0.45\linewidth]{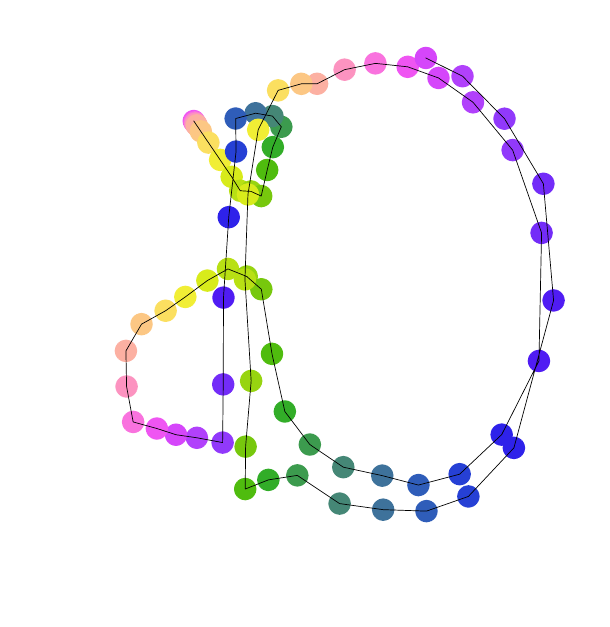}}
  \\
\subfloat[Urban sounds (SONYC-UST excluding speech and music).]{
  \includegraphics[width=0.45\linewidth]{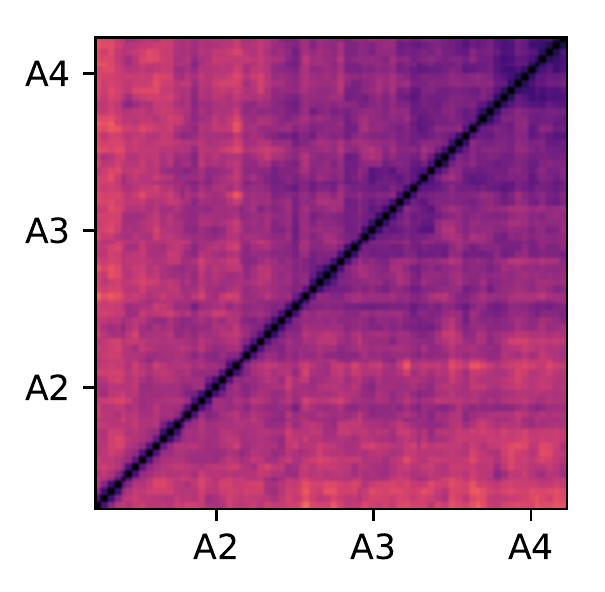}
  \includegraphics[width=0.45\linewidth]{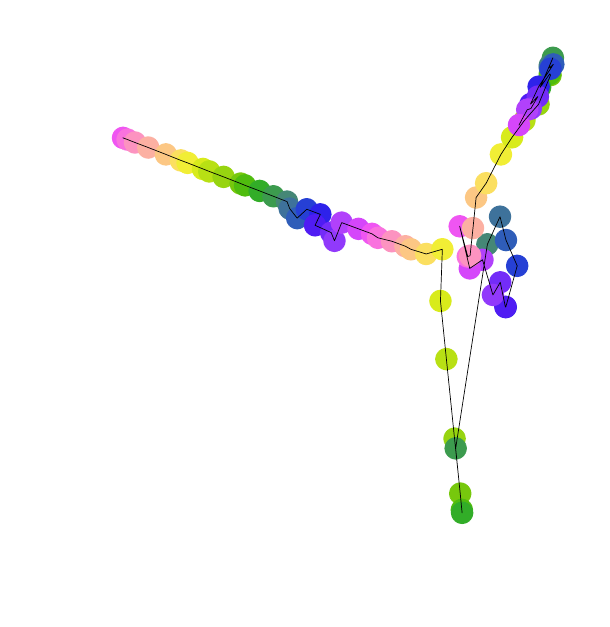}}
  \caption{Extension of the proposed method to speech (a) and environmental soundscapes (b).
  In both cases, the number of neighbors is set to $K=3$ and the loudness mapping $\Lambda$ is logarithmic.
  See Section \ref{sec:experiments-nonmusic} for details.}
  \label{fig:non-music-data}
\end{figure}

We analyze a portion of the SONYC Urban Sound Tagging dataset (SONYC-UST v0.4), which contains a collection of 3068 acoustic scenes from a network of autonomous sensors in various locations of New York City \cite{cartwright2019sonyc}.
We restrict this collection to the acoustic scenes in which the consensus of expert annotators has confirmed the absence of both human speech and music.
As a result of this preliminary curation step, we obtain $N=233$ audio recordings from eight different sensor locations.
Each of these scenes contains one or several sources of urban noise pollution, among which: engines, machinery and non-machinery impacts, powered saws, alert signals, and dog barks.
Figure \ref{fig:non-music-data} (b) shows that no discernible correlations across octaves are observed in this dataset.
This finding confirms the conclusions of a previous publication \cite{muller2011jstsp}, which stated that ``music audio signal processing techniques must be informed by a deep and thorough insight into the nature of music itself''.

\section{Conclusion}
\label{sec:conclusion}

The Isomap manifold learning algorithm offers an approximate visualization of nearest neighbor relationships in any non-Euclidean metric space.
Thus, a previous publication \cite{leroux2008neurips} proposed to apply Isomap onto cross-correlations between grayscale intensities in natural images.
In this article, we have borrowed from the protocol of this publication to apply it on music data.

Despite their methodological resemblance, the two studies lead to different insights.
While \cite{leroux2008neurips} recovered a quasi-uniform raster, we do not recover a straight line from cross-correlations along the log-frequency axis.
Instead, we obtain a cylindrical lattice in dimension three.
Assigning pitch classes to  subbands reveals that this lattice is akin to a Drobisch-Shepard helix \cite{shepard1964jasa}, which makes a full turn at each octave.
Thus, whereas \cite{leroux2008neurips} learned a 2-D topology from 2-D data, we learned a 3-D topology from 1-D data.
Furthermore, after benchmarking several design choices, we deduce that the most regular helical shape results from: a diverse instrumentarium; a graph of $K=3$ nearest neighbors; and a logarithmic loudness mapping.
Lastly, we have discussed the limitations of our findings: although spoken vowels also exhibit a quasi-helical topology in subband neighborhoods, the same cannot be said of urban noise.

Beyond the realm of manifold learning, the present article motivates the development of weight sharing architectures that foster octave equivalence in deep representations of music data.
Three examples of such architectures are:
spiral scattering transform \cite{lostanlen2015dafx};
spiral convolutional networks \cite{lostanlen2016ismir}; and
harmonic constant-$Q$ transform \cite{bittner2017ismir}.
Future work will extend this protocol to bioacoustic data, and comparing the influence of species-specific vocalizations onto the empirical topology of the frequency domain.

%

%

\section{Acknowledgment}
The authors wish to thank St\'{e}phane Mallat for helpful discussions.

\label{sec:refs}

\bibliographystyle{IEEEbib}
\bibliography{refs}

\end{document}